\documentclass[prl,showpacs,showkeys,twocolumn,floatfix]{revtex4}

\usepackage{graphicx}
\usepackage{units}
\usepackage{xspace}
\usepackage[american]{babel}
\usepackage{color}

\begin{document}


\title{Observation of dipole-dipole interaction in a degenerate quantum gas}

\author{J.\ Stuhler}
\email{j.stuhler@physik.uni-stuttgart.de}
\author{A.\ Griesmaier}
\author{T.\ Koch}
\author{M.\ Fattori}
\author{T.\ Pfau}

\affiliation{5.\ Physikalisches Institut, Universit\"at Stuttgart,
Pfaffenwaldring 57, 70550 Stuttgart, Germany}

\author{S.\ Giovanazzi}
\affiliation{Physikalisches Institut, Universit\"at T\"ubingen,
Auf der Morgenstelle 14, 72076 T\"ubingen, Germany}

\author{P.\ Pedri}
\altaffiliation[present address: ]{LPTMS, B\^{a}timent 100,
Universit\'{e} Paris-Sud, 91405 ORSAY cedex, France}
\author{L.\ Santos}
\affiliation{Institut f\"ur Theoretische Physik III, Universit\"at
Stuttgart, Pfaffenwaldring 57, 70550 Stuttgart, Germany}

\date{\today}

\begin{abstract}
We have investigated the expansion of a Bose-Einstein condensate
(BEC) of strongly magnetic chromium atoms. The long-range and
anisotropic magnetic dipole-dipole interaction leads to an
anisotropic deformation of the expanding Cr-BEC which depends on
the orientation of the atomic dipole moments. Our measurements are
consistent with the theory of dipolar quantum gases and show that
a Cr-BEC is an excellent model system to study dipolar
interactions in such gases.
\end{abstract}

\maketitle

Ultracold or even degenerate atomic quantum gases are typically
very dilute systems. Nevertheless, interatomic interactions
strongly determine many of the observed phenomena and their
underlying physics \cite{BEC_review2,BEC_review}. Until recently,
only short-range and isotropic interactions have been considered.
However, recent developments in the manipulation of cold atoms and
molecules are paving the way towards the analysis of polar gases
in which dipole-dipole interparticle interactions are important.
In this sense, exciting perspectives are opened by recent
experiments
\cite{doyle:2004,Sage:2005,Stan:2004,Inouye:2004,Simoni:2003b,Meerakker:2005}
on cooling and trapping of polar molecules as well as on
photoassociation and on Feshbach resonances in binary mixtures of
ultracold atoms. So far, these methods did not provide a sample of
polar molecules cooled to or produced in the degeneracy regime.

However, a dipolar quantum gas can also be obtained exploiting the
large magnetic moment of some atomic species. We have recently
realized such a dipolar degenerate quantum gas by creating a BEC
of $^{52}$Cr atoms~\cite{Griesmaier:2005a}. Chromium atoms possess
very large magnetic dipole moments of $6$ Bohr magnetons. As a
consequence, the magnetic dipole-dipole interaction (MDDI) is much
stronger than in previously realized BECs, e.g. by a factor of
$36$ compared to alkali atoms. Hence, magnetic dipole-dipole
forces between the particles start to play an important role.

New exciting phenomena are expected in dipolar quantum gases
oriented by an external field since the particles interact via
dipole-dipole interactions which are long-range and anisotropic.
Recent theoretical analyses have shown that stability and
excitations of dipolar gases are crucially determined by the trap
geometry
\cite{Goral:2000b,Yi:2000a,Santos:2000a,Yi:2002,Goral:2002,Baranov:2002b,ODell:2004a,Santos_Roton03}.
Dipolar degenerate quantum gases are also attractive in the
context of strongly-correlated atoms
\cite{Goral:2002a,Baranov:2005,Rezayi:2005}, as physical
implementation of quantum computation \cite{DeMille:2002a}, and
for the study of ultracold chemistry \cite{Bodo:2002}.

It has also been predicted that the dipole-dipole interaction
modifies the condensate shape in a trap
\cite{Goral:2000b,Yi:2000a,Santos:2000a} and during the expansion
after release from a trap~\cite{Giovanazzi:2003a}. In particular,
a homogeneous magnetic field is expected to align the magnetic
dipoles of the atoms in a Cr-BEC and the MDDI between them leads
to an anisotropic change of the shape of the gas. This effect may
be considered as dipole-dipole induced magnetostriction.
Magnetostriction was discovered by J. P. Joule more than 150 years
ago when observing the deformation of an iron bar exposed to a
magnetic field~\cite{Joule:1847}. Since then, it has been
extensively studied experimentally in magnetic
solids~\cite{Gibbs:2000} and
liquids~\cite{Huang:2004,Odenbach:2004}. Especially classical
dipolar fluids have attracted much attention over the past
years~\cite{Depeyrot:2001,Groh:1999}. In the general context of
effects of homogeneous external fields on gases, it's worthwhile
to mention Senftleben-Beenakker effects which are correlated with
transport properties of molecules~\cite{Beenakker:1970} but not
based on dipole-dipole interactions.

So far, dipole-dipole interactions in gases were typically much
smaller than other interactions or the kinetic energy of the
particles (temperature). Consequently, to our knowledge,
mechanical effects of dipole-dipole interaction have not been
observed in gaseous systems. The situation is crucially different
in a degenerate quantum gas of chromium atoms because of its
extremely low temperature in the nK range and the large magnetic
dipole moment of the atoms.

In this Letter, we report on the observation of magnetic
dipole-dipole interaction in a degenerate quantum gas. We
investigate the expansion of a chromium Bose-Einstein condensate
polarized by a homogeneous magnetic field and show that the
long-range and anisotropic character of the dipole-dipole
interaction leads to an anisotropic deformation of the expanding
Cr-BEC. This manifestation of dipole-dipole interaction opens
exciting perspectives for the analysis of other interesting
dipole-induced phenomena, as those discussed above. In the
following, we first briefly describe the experimental procedure we
used to obtain the expansion data which are presented
subsequently. In order to provide a basic understanding of the
underlying physics, we then explain qualitatively the mechanisms
that are responsible for the reported observations. This is
followed by a rigorous theoretical treatment of the Cr-BEC
expansion within the framework of dipolar superfluids. The theory
is obtained without free parameters and agrees very well with the
experimental data.

Our experimental investigation of dipolar effects in a degenerate
quantum gas starts with the production of a Cr-BEC. As described
in Ref. \cite{Griesmaier:2005b}, this requires novel cooling
strategies that are adapted to the special electronic and magnetic
properties of chromium atoms. The final step to reach quantum
degeneracy is forced evaporative cooling within a crossed optical
dipole trap. We observe Bose-Einstein condensation at a critical
temperature of $ T_{\rm c}\sim 700$~nK. At $T\ll T_{\rm c}$ almost
pure condensates with up to 100,000 $^{52}$Cr atoms remain.

To measure the influence of the magnetic dipole-dipole interaction
on the condensate expansion, we prepare a $^{52}$Cr-BEC in the
crossed optical dipole trap. In the BEC, the atoms are fully
polarized in the energetically lowest Zeeman substate ($m_J=-3$).
We then adiabatically change the laser intensities to form a trap
with frequencies of $\omega_x/2\pi = 942(6)$~Hz, $\omega_y/2\pi =
712(4)$~Hz and $\omega_z/2\pi = 128(7)$~Hz. This results in
elongated trapped condensates oriented along the $z$-axis. A
homogeneous magnetic offset field of $B=|\vec{B}| \sim 1.2$~mT
defines the orientation of the atomic magnetic dipole moments, the
direction of magnetization. $\vec{B}$ is either kept along the
$y$-direction for transversal magnetization or slowly (within 40
ms) rotated to the $z$-direction for longitudinal magnetization.
After a holding time of 7 ms, the atoms are released from the trap
by switching off both laser beams. The condensate expands freely
for a variable time, is subsequently illuminated with a resonant
laser beam and its shadow is recorded by a calibrated CCD camera.
We determine the relevant parameters, like BEC atom number and BEC
sizes, by two-dimensional fits of parabolic function to the
resulting absorption image.

\begin{figure}[hhh!]
 \includegraphics[width=0.49\textwidth]{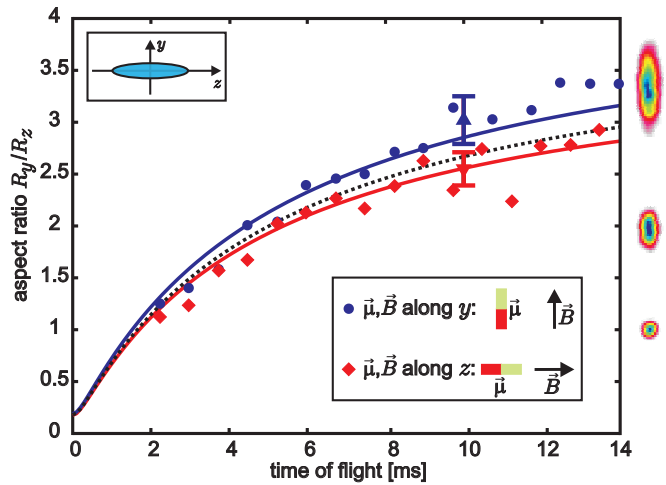}
  \caption{\label{figure1}Aspect ratio of a freely expanding Cr-BEC
for two different directions of magnetization induced by a
homogeneous magnetic field ($\vec{B}$). Blue: Experimental data
(circles) and theoretical prediction without adjustable parameter
(solid line) for transversal magnetization (atomic dipoles
$\vec{\mu}$ aligned orthogonal to the weak trap axis). Red: Data
(diamonds) and theory curve (solid line) for longitudinal
magnetization ($\vec{\mu}$ parallel to the weak trap axis). Blue
upward and red downward triangles are results of 31 measurements
taken 10 ms after release for transversal and longitudinal
magnetization, respectively. Dashed line: theory curve without
dipole-dipole interaction. The inset (upper left corner) sketches
the in-trap BEC. The BEC images at the right axis illustrate the
condensate shape for some aspect ratios.}
\end{figure}

A convenient measurement quantity for the expansion is the aspect
ratio of the condensate. In our experiment, it is defined as
$R_{y}/R_{z}$, the BEC extension along one axis of strong
confinement divided by the extension along the weak axis of the
trap. This quantity is not very sensitive to the exact number of
atoms but only to the trap geometry and the ratio between the MDDI
and the short-range interaction. Figure~\ref{figure1} shows the
aspect ratio for different times of free expansion. As indicated
by the theory curve for vanishing dipole-dipole interaction
(dashed line), a non-dipolar BEC would expand with an inversion of
the aspect ratio~\cite{BEC_review}. The MDDI leads to significant
deviations from the non-dipolar behavior altering the aspect-ratio
subject to the direction of magnetization. The experimental data
points correspond to two different directions of magnetization,
transversal (blue circles) and longitudinal (red diamonds) with
respect to the weak trap axis. The data clearly reveal the
influence of MDDI since changing from transversal to longitudinal
magnetization leads to a decrease in the aspect ratio. The
triangles with statistical error bars at a fixed expansion time of
10 ms represent the results of 31 measurements each and prove the
significance of the observed dipole-dipole induced effect.

\begin{figure}[hhh!]
 \includegraphics[width=0.35\textwidth]{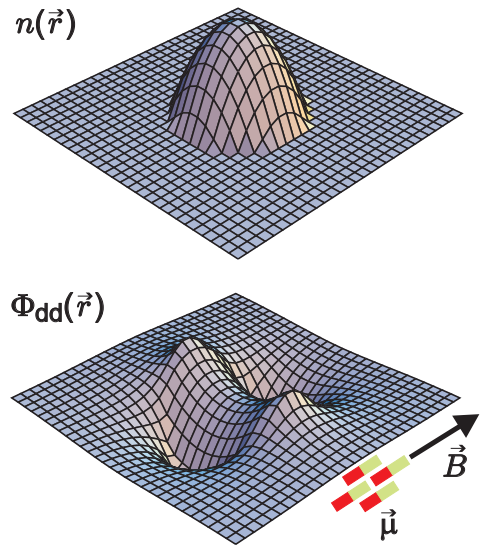}
  \caption{\label{figure2}Top: Sketch of the atomic density distribution $n(\vec{r})$ of a non-dipolar BEC in the
  Thomas-Fermi regime within a spherically symmetric harmonic trap ($x,z$-plane cross-section through
  the center of trap). Bottom: Asymmetric dipole-dipole interaction potential $\Phi_{\rm dd}(\vec{r})$
  (cross-section like top) for a dipolar BEC with density distribution $n(\vec{r})$.
  Within the atomic cloud, $\Phi_{\rm dd}(\vec{r})$ has the form of a saddle with negative curvature
along the
  direction of magnetization (sketched by the magnets with dipole moment $\vec{\mu}$ and the $\vec{B}$ vector) and positive curvature orthogonal to it.}
\end{figure}

The modification of the aspect ratio is explained by the
anisotropic nature of the MDDI, which leads to an anisotropic
deformation of the condensate. More precisely, the BEC is
stretched along the magnetization direction and squeezed
orthogonal to it. Consequently, transversal magnetization
increases the aspect ratio and longitudinal magnetization
decreases it. The elongation of the BEC along the direction where
two individual dipoles attract each other and its compression in
the direction where they repel each other seems at first sight
counter-intuitive. However, this apparent paradox can be
understood by considering the total energy of a trapped BEC. For
simplicity, we consider here a spherically symmetric harmonic
trapping potential but similar considerations also hold for
non-axisymmetric traps. In the Thomas-Fermi regime, where the
quantum pressure is negligibly small compared to the interaction
energy stemming from the short-range and isotropic inter-particle
interaction of mainly van der Waals type (so-called contact
interaction), the atomic density distribution $n(\vec{r})$ of a
non-dipolar BEC has the form of a spherically symmetric inverted
paraboloid (figure \ref{figure2}, top). The meanfield potential
arising from the contact interaction is proportional to
$n(\vec{r})$ and hence directly reflects the atomic density
distribution.

In a dipolar BEC, however, the MDDI breaks the symmetry and leads
to an additional dipole-dipole potential $\Phi_{\rm dd}(\vec{r})$.
In contrast to the meanfield potential, $\Phi_{\rm dd}(\vec{r})$ -
due to the long-range character of the MDDI - is not proportional
to the local density but has to be evaluated by integrating over
the whole atomic density distribution. It can be
shown~\cite{ODell:2004a,Eberlein:2005} that the resulting
dipole-dipole potential within the cloud of atoms has the
(parabolic) form of a saddle with negative curvature along the
magnetization direction and positive curvature orthogonal to it
(figure \ref{figure2}, bottom). As a consequence, the total energy
of a trapped BEC is lowered if the atoms are redistributed from
the repulsive to the attractive direction, even at the cost of
increasing the external trapping potential energy. At this point,
we would like to note that the repulsive contact interaction
stabilizes the BEC against collapse by keeping the atoms at
distance, and hence more atoms along a direction result in an
overall elongation of the condensate in this direction.
After switching off the trapping potential, there remain only two
contributions to the potential energy. One part, the meanfield
potential due to the contact interaction, has the form of an
inverted paraboloid. The other part is $\Phi_{\rm dd}(\vec{r})$
which due to its saddle-like shape leads to an increased expansion
of the BEC along the direction of magnetization and to a decreased
expansion orthogonal to it if compared to a non-dipolar BEC of
same shape. Hence, the general trend of the BEC reshaping -
elongation along the magnetization and contraction orthogonal to
it - is conserved also during the
expansion~\cite{Giovanazzi:2003a}.

A complete quantitative theoretical understanding of the expansion
of the Cr-BEC can be obtained by self-similar Thomas-Fermi
solutions of the superfluid hydrodynamic equations for a dipolar
condensate~\cite{ODell:2004a}. We initially calculate the
Thomas-Fermi density profile $n_0(\vec{r})$ before opening the
trap by solving the equation
\begin{equation}\label{equation1}
\mu = V_{\rm ext}(\vec{r}) + g n_0(\vec{r}) + \int d^3r' U_{\rm
dd}(\vec{r}-\vec{r}\,') n_0(\vec{r}\,')\;\;.
\label{TF}
\end{equation}
In Eq.~(\ref{equation1}), $\mu$ is the chemical potential, and
 $V_{\rm ext}(\vec{r})=\frac{\textstyle m}{\textstyle
2}\left(\omega_x^2x^2+\omega_y^2 y^2+\omega_z^2 z^2\right)$ is the
trapping potential with frequencies $\omega_{x,y,z}$ and atomic
mass $m$. $g=4\pi \hbar^2 a/m$, with $s$-wave scattering length
$a$, is the mean-field coupling constant of contact-like atom-atom
interactions.
\begin{equation}\label{udd} U_{\rm
dd}(\vec{r})=\frac{\textstyle \mu_0 \mu^2_{\rm m}}{\textstyle 4
\pi r^3} \left( 1- \frac{\textstyle 3
(\hat{e}_\mu\vec{r})^2}{\textstyle r^2}\right)
\end{equation}
is the interaction energy between two magnetic dipoles
($\vec{\mu}_{\rm m}=\mu_{\rm m} \hat{e}_\mu$, $\mu_{\rm
m}=6\mu_{\rm B}$)
aligned by a polarizing magnetic field
($\hat{e}_\mu \| \vec{B}$) and with relative coordinate $\vec{r}$.

In spite of the complicated non-local form of Eq.~(\ref{TF}), the
density profile remains an inverted paraboloid~\cite{ODell:2004a}
of the form $n_0(\vec{r})=n_{00}\left(1-(x/R_x)^2
-(y/R_y)^2-(z/R_z)^2\right) $, where $R_{x,y,z}$ are the
Thomas-Fermi radii, as for the case of short-range interacting
condensates.

The expansion dynamics is then given~\cite{ODell:2004a} by a
self-similar solution~\cite{Castin1996a,Kagan1997b} for the
density
\begin{equation}\label{densityselfsim}
n(\vec{r},t)=n_0\left( \{r_i/b_i(t) \} \right)/\prod_i b_i(t)
\end{equation}
($i=x,y,z$ labels the spatial coordinates) and the hydrodynamic
velocity field $\vec{v}(\vec{r},t)=\{ v_i\}$ with components
$v_i=\dot{b}_i(t) r_i/b_i(t)$. In case of non-axisymmetric traps,
both the equation for the radii of the condensate in equilibrium
and the equations for evolution of the scaling parameters $b_i$
during the expansion dynamics involve long analytical expressions
that will be reported elsewhere.

The solid lines in figure~\ref{figure1} represent the
corresponding theoretical predictions obtained without any free
adjustable parameters. Only measured or known quantities, namely
atom number, trap frequencies, $s$-wave scattering length that
characterizes the repulsive contact interaction~\cite{Werner:2005}
and magnetic moment have been included. Compared to the
calculations for negligible dipole-dipole interaction (dashed
line), the aspect ratio of the expanding Cr-BEC is - in good
agreement with the experimental results - increased for
transversal magnetization and decreased for longitudinal
magnetization.

Summarizing, the described experiments with a chromium BEC
constitute the observation of magnetic dipole-dipole interaction
in a quantum gas, which is, to the best of our knowledge, the
first mechanical manifestation of dipole-dipole interaction in a
gas. In particular, applying a homogeneous magnetic field which
magnetizes the Cr-BEC leads to a redistribution of the trapped
atoms. Similar to what is known from magnetic solid particles or
liquids (ferrofluids), the strongly magnetic chromium atoms align
preferably along the direction of magnetization under the
influence of the magnetic field. This MDDI induced change in shape
remains visible also after release of the Cr-BEC from the trap.
The expansion of the Cr-BEC is well described within the framework
of dipolar superfluids. The experimental results prove that the
MDDI in a Cr-BEC is strong enough to significantly influence the
condensate properties and to lead to measurable effects. In this
sense, a Cr-BEC opens fascinating perspectives for the
experimental study of dipole-dipole interaction induced magnetism
in gaseous systems. Since one can exploit Feshbach
resonances~\cite{Werner:2005} to adjust contact-like (isotropic
and short-range) atom-atom interactions and use rotating magnetic
fields to tune the dipole-dipole
interaction~\cite{Giovanazzi:2002a}, interaction regimes ranging
from only contact to purely dipolar can be realized. Depending on
the relative strengths of these two interactions and on the
absolute strength of the dipole-dipole interaction, many exciting
phenomena are expected.
\begin{acknowledgments}
We thank our atom optics group for encouragement and practical
help. This work was supported by the Alexander von Humboldt
Foundation and the German Science Foundation (DFG) (SPP1116 and
SFB/TR 21).
\end{acknowledgments}

\bibliography{crbib}

\end{document}